\def\lam{\lambda}
\def\1214{FBQS~1214+2803}
\def\kms{km~s$^{-1}$}

\documentclass[12pt,preprint]{aastex}

\shortauthors{Branch et al.}
\shorttitle{Spectrum of the FeLoBAL Quasar \1214}

\begin{document}
\title {The Spectrum of the FeLoBAL Quasar \1214: A Resonance--Scattering 
Interpretation}

\author {David Branch, Karen M. Leighly, R. C. Thomas, and E.~Baron}

\affil{Department of Physics and Astronomy, University of
Oklahoma, Norman, Oklahoma 73019, USA}

\begin{abstract}

The usual interpretation of the spectrum of a BALQSO is that a
broad--band continuum from the central engine plus broad emission
lines from a surrounding region (the BELR) emerges from near the
center of the QSO, and broad absorption lines are superimposed in a
separate outlying region (the BALR).  For \1214, designated as an
FeLoBAL QSO because its spectrum contains numerous absorption features
from excited states of Fe~II, we explore an alternative interpretation
based on resonance scattering.  In this model line emission and
absorption occur in the same line--forming region (the LFR).  A
resonance--scattering synthetic spectrum computed with the
parameterized supernova synthetic--spectrum code SYNOW fits the
spectrum of \1214 rather well, so the resonance--scattering model
merits further study.  Some implications of the model and its possible
applicability to other QSOs are briefly discussed.
      
\end{abstract}


\section{INTRODUCTION}

BALQSOs --- quasars that have broad absorption lines in their spectra
--- can be divided into LoBALs and HiBALs, which do and do not have
strong absorption lines produced by low--ionization species such as
Mg~II, Al~III, and Fe~II.  LoBALs that have numerous lines from
excited states of Fe~II are called FeLoBALs (Becker et~al. 1997,
2000).

The usual interpretation of the spectrum of an FeLoBAL QSO is similar
to that of the spectra of other BALQSOs.  The effective continuous
spectrum consists of a true continuum from the accretion disk, plus
broad emission lines (BELs) that form in a surrounding region --- the
BELR.  Broad absorption lines (BALs) are superimposed on the effective
continuum as it passes through a separate region --- the BALR.   The
distance of the BALR from the QSO center may be large compared to the
size of the BELR, and the global covering factor --- the fraction of
the sky covered by the BALR as viewed from the center of the QSO ---
may be small.

In this {\sl Letter} we explore an alternative interpretation of the
spectrum of a particular FeLoBAL QSO.  In our simple model, the
rest--frame UV spectrum consists of P~Cygni features formed by
resonance scattering, superimposed on a continuum.  The line emission
and absorption components come from the same spherically--symmetric
line--forming region --- the LFR.  To illustrate this interpretation
we concentrate on an FeLoBAL that was discovered in the First Bright
Quasar Survey (White et~al. 2000) and is designated FIRST
J121442.3+280329, or
\1214 for short.  We choose this particular FeLoBAL because it has
recently been analyzed in detail, in the context of the usual BAL
model, by de~Kool et~al. (2002; hereafter dK02), and because its
spectrum, although rich in lines, appears to be amenable to an
analysis based on a single LFR.  In \S2 we describe the modeling of
\1214 by dK02.  The alternative resonance--scattering interpretation
is presented in
\S3. Some of the implications of the resonance--scattering 
model, and its possible applicability to other QSOs, are briefly
discussed in \S4.

\section{A RECENT INTERPRETATION OF THE SPECTRUM OF \1214}

Because we make use of the results of dK02, and we want to compare and
contrast the results of our model and theirs, we must first briefly
describe their analysis.  They studied a spectrum obtained on 1998
May~18 using the High Resolution Echelle Spectrometer (HIRES; Vogt
et~al. 1994) on the Keck~I 10~m telescope.  The redshift was
determined to be $z=0.692$.

In their analysis dK02 used effective continuous spectra that
consisted of a power--law continuum plus Fe~II and Mg~II BELs.  The
Mg~II BEL was the sum of two Gaussians centered on the two components
of the Mg~II $\lam2798$ doublet ($\lam\lam$2796, 2803).  Two different
templates for the Fe~II BELs were considered.  The first consisted of
a linear combination of five sets of Fe~II BELs from theoretical model
calculations (Verner et~al. 1999), and the second was the observed
Fe~II BEL spectrum of the strong emission--line QSO 2226--3905
(Graham, Clowes, \& Campusano 1996).

For the absorption features, dK02 obtained a template distribution of
line optical depth with respect to velocity in the BALR from the
observed absorption profile of Fe~II $\lam3004$, an apparently
unblended line of moderate strength.  Given the assumption that only
absorption takes place in the BALR, the optical depth was obtained
from $\tau(v) = -ln\ F_\lam$, where $F_\lam$ is the fractional
residual flux in the absorption feature.  The resulting optical--depth
distribution extended from about 1200 to 2700~\kms\ and peaked near
2100~\kms.  This optical--depth distribution, scaled in amplitude, was
used for all absorption lines.  For each of the absorbing ions that
were introduced --- Fe~II, Mg~II, Cr~II, and Mn~II --- the column
density was a fitting parameter.  The relative strengths of the lines
of each ion turned out to be consistent with LTE.

Three models that differed in their details (see dK02) were presented,
with similar results. The column densities of Fe~II, Cr~II, and Mn~II
were well constrained.  (The column density of Mg~II could not be well
constrained because the only Mg~II absorption, due to $\lam2798$, is
saturated.)  The excitation temperature was found to be near 10,000~K.
Two local covering factors, representing the fractions of the
power--law source and the BELR that are covered by the BALR as viewed
by the observer, were introduced to reproduce the observed
``non--black saturation'' --- the fact that in the observed spectrum
even very strong absorption features do not go to zero flux.  Both
local covering factors were found to be $0.7\pm0.1$.

A detailed view of the spectral fit for one of the models was
presented, and practically all of the observed absorptions were
reasonably well accounted for.  To further interpret the results of
their spectrum fits dK02 then used the photoionization--equilibrium
code CLOUDY (Ferland 2000) to compute a grid of constant--density slab
models irradiated by a range of ionizing spectra.  The ionization
parameter $U$, the hydrogen density $n_H$, and the hydrogen column
density $N_H$ were found to satisfy $-2.0 < log\ U < -0.7$, $7.5 <
log\ n_H < 9.5$, and $21.4 < log\ N_H < 22.2$.  From these values the
distance of the BALR from the center of the QSO was inferred to be
between 1 and 30~pc.

\section{A RESONANCE--SCATTERING INTERPRETATION}

We explore a resonance-scattering interpretation using synthetic
spectra generated with the fast, parameterized synthetic--spectrum
code SYNOW, which often is used for making line identifications and
initial coarse analyses of photospheric--phase supernova spectra.  The
use of SYNOW, as well as more elaborate and physically
self--consistent spectrum--synthesis codes such as PHOENIX, for the
analysis of supernova spectra is reviewed by Branch, Baron, \& Jeffery
(2002).  SYNOW assumes spherical symmetry and a sharp photosphere that
emits a blackbody continuum characterized by temperature $T_{bb}$.
Expansion velocity is proportional to radius, as expected for matter
that has been coasting at constant velocity after an impulsive
ejection with a range of velocities.  Line formation, treated in the
Sobolev approximation, occurs by resonance scattering of photons from
the photosphere.  Line blending (multiple scattering) is treated
exactly, within the context of the Sobolev approximation.  Line
optical depths are taken to decrease with radius according to a power
law.  For each ion whose lines are introduced, the optical depth of a
``reference line'' at the inner boundary of the LFR is a fitting
parameter, and the optical depths of the other lines of the ion are
determined by assuming LTE level populations at excitation temperature
$T_{exc}$.  Oscillator strengths are from Kurucz (1993).  In
supernovae, line formation ordinarily takes place immediately above
the photosphere, i.e., the bottom of the LFR has the velocity at the
photosphere, $v_{phot}$, but sometimes it is appropriate to ``detach''
the lines of an ion by allowing line formation only above some
velocity $v_{min}$ that exceeds $v_{phot}$.  Individual values of
$v_{min}$, and also of $v_{max}$, can be assigned to each ion.

For \1214\ we use $T_{bb} = 8000$~K to approximate the shape of the
underlying continuum, we take $T_{exc} = 10,000$~K from the analysis
of dK02, and we introduce only lines of Mg~II, Fe~II, and Cr~II.  (In
the synthetic spectrum of dK02 the Mn~II lines play only a minor
role.)  The index of the optical--depth power law is taken to be $n=2$
for all ions.

Fig.~1 compares the strong observed feature produced by Mg~II
$\lam2798$ with a synthetic spectrum that contains only lines of Mg~II
and has $v_{phot}=1000$ \kms, $v_{max}= 2800$~\kms.  The synthetic
P~Cygni profile is so saturated that it not sensitive to the optical
depth distribution, only to the values of $v_{phot}$ and $v_{max}$.
It is interesting that the observed feature can be fit this well
with the simple assumption of resonance scattering in a spherically
symmetric LFR.  Note that no local covering factor has had to be
introduced; because line emission and absorption take place in the
same LFR, the saturated absorption feature does not go black.

Fig.~2 compares the spectrum of \1214 with a synthetic spectrum that
contains lines of Mg~II, Fe~II, and Cr~II, all with $v_{max}=2800$
\kms.  In order to make
their absorption components blueshifted enough, the Fe~II and Cr~II
lines have $v_{min}=1800$ \kms.  Cr~II and Fe~II contribute some lines
of weak and moderate strength to the right of the Mg~II feature, while
Fe~II lines dominate the spectrum to the left of the Mg~II feature.
The overall fit is good, although there are some discrepancies.  The
fact that the synthetic continuum goes too high at the red end of the
spectrum is not a serious concern because there is no reason that the
true continuum should be that of a blackbody.  The strong Fe~II blends
near 2740, 2620, and 2400~\AA\ are reasonably well matched, although
now that the Mg~II feature is blended with numerous Fe~II lines it
does not fit as well as in Fig.~1.  The Cr~II absorption to the right
of the Mg~II emission is too strong.  A closer look at a crowded part
of the spectrum (Fig.~3) shows that most of the details are fit rather
well, considering the simplicity of our model and especially that we
are using a power--law optical--depth distribution rather than a
customized (template) distribution of optical depth with respect to
velocity.  Why do the Fe~II and Cr~II features form only above 1800
\kms\ while Mg~II forms above 1000~\kms?  The simplest possibility is
that the true optical--depth distributions of Mg~II, Fe~II, and Cr~II
lines with respect to velocity are of similar shape, but their
relative amplitudes are such that the optical depths of the Fe~II and
Cr~II lines are less than unity below 1800 \kms\ while the optical
depth of the very strong Mg~II resonance feature is well above unity.

An extension of the synthetic spectrum of Fig~2. to longer wavelengths
is practically featureless.  For anything like solar abundances, the
only lines in the rest--frame optical spectrum that necessarily should
be expected in this model are the hydrogen lines, although some lines
of other ions that we have not used for the UV spectrum, such as the
H\&K lines of Ca~II, might have significant optical depths.  An
extension of the synthetic spectrum of Fig.~2 to wavelengths shorter
than 2300~\AA\ has numerous features of moderate strength, and strong
features also are to be expected from ions such as Al~III and Al~II
that are not needed for Fig.~2.

\section{DISCUSSION}

The resonance--scattering model can fit the spectrum of \1214 rather
well with simple assumptions, and if we were to introduce a customized
distribution of optical depth with respect to velocity the fit
certainly could be improved.  Nevertheless, the model is at best an
idealization, and it needs more study before we can be sure that it is
applicable.  Our use of an $n=2$ power law for the optical depths is
just illustrative because here a power--law has no particular
justification; the 8000~K blackbody continuum is just a surrogate for
the real continuum from the central source; and our assumption that
velocity is proportional to distance from the center may not be
correct.  The most important feature of this model is that the red
side of the emission component comes from the far side of the QSO
rather from a central BELR, so it would mean that the LFR is more or
less spherically symmetric and continuous and the global covering
factor is large.  Since emission and absorption components come from
the same LFR, the fact that the same lines are observed in emission
and absorption would necessarily follow.  Local covering factors would
not necessarily be required to account for non--black saturation.  For
lines of more than moderate strength, column densities inferred from
the resonance scattering intepretation would be higher than those
inferred from the standard model because with the presence of emission
in the LFR, larger optical depths would be required to match a given
absorption depth in the observed spectrum.  Thus the excitation
temperature inferred from a detailed line by line study could change.

Taking $M_B=-26.3$ from Becker et~al. (2000) and assuming a standard
AGN spectral energy distribution (Mathews \& Ferland 1987) we estimate
the luminosity of \1214 to be roughly $6 \times 10^{46}$ erg~s$^{-1}$.
If, for example, the luminosity is ten percent of the Eddington
luminosity then the black--hole mass is $5\times10^9$~M$_\odot$, and
the escape velocity falls to 1000 \kms\ at a radius near 20~pc.  If
the inner radius of the LFR is at least this large, then the LFR does
not contain a true photosphere, because a blackbody having $T=8000$~K
and $L= 6\times10^{46}$ erg~s$^{-1}$ would have a radius of only $1.4
\times 10^{17}$ cm, or 0.045~pc.  The column density of singly ionized
iron required to produce the weak Fe~II lines is $10^{17}$ cm$^{-2}$
(dK02), which with a solar iron abundance and {\sl all} iron being
singly ionized corresponds to a bare minimum hydrogen column density
of $10^{21.4}$ cm$^{-2}$.  On the other hand, if the LFR is optically
thick to electron scattering the column density exceeds $10^{24}$
cm$^{-2}$.  Then the mass of the LFR could be large, $6\times10^7
r_{20}^2 N_{24}$~M$_\odot$, where $r_{20}$ is the inner radius of the
shell in units of 20~pc and $N_{24}$ is the column density in units of
$10^{24}$ cm$^{-2}$.

The kinetic--energy luminosity of $\sim10^{45} r_{20} N_{24}$
erg~s$^{-1}$ is not necessarily large compared to the QSO luminosity
but the mass flow rate of $\sim3000 r_{20} N_{24}$~M$_\odot$~y$^{-1}$
may be large compared to the accretion rate required to provide the
QSO luminosity ($\sim10~$M$_\odot$~y$^{-1}$, if the efficiency of
conversion of accretion energy to radiation is ten percent), in which
case the LFR is a consequence not of a steady--state wind but an
episodic ejection.

Resonance scattering in differentially expanding LFRs having high
global covering factors was explored years ago (e.g., Scargle, Caroff,
\& Noerdlinger 1970; Surdej \& Swings 1981; Drew \& Giddings 1982,
Surdej \& Hutsem\'ekers 1987) but it fell out of favor as a model for
BALQSOs in general, for various reasons.  Weymann et al. (1991)
found that HiBALs and non--BALQSOS have similar emission lines,
suggestive of a single population of QSOs having a small global
covering factor, with HiBALs being seen from a special orientation.
Emission--line equivalent widths often are too small to account for
the photons scattered out of the deep absorption troughs (barring some
photon destruction mechanism, e.g., absorption by dust); from analysis
of emission and absorption line profiles Hamann, Korista
\& Morris (1993) concluded that high global covering fractions were not
favored for the HiBALs of the Weymann et~al. sample.  However, Becker
et~al. (2000) and Gregg et~al. (2002) have argued against the usual
scenario that BALQSOs are normal quasars seen edge--on, and in favor
of an alternative picture in which BALQSOs are an early stage in the
development of new or refueled quasars.  (Evolution, not orientation.)
  
We suggest that the resonance--scattering model should be
reconsidered, for the FeLoBAL QSOs.  Our preliminary modeling of the
spectra of FeLoBALs other than \1214 indicates that the
resonant--scattering model is able to account for the spectra of at
least some of them.

Although the model clearly is not applicable to all BALQSOs, it may
have some relevance to LoBAL QSOs in general.  LoBALs have different
emission--line properties from HiBALs, including a small equivalent
width of [O~III] $\lambda5007$ (Boroson \& Meyers 1992).  The [O III]
line is produced far enough from the center of the QSO that its
emission should be isotropic (Kuraszkiewicz et al. 2000), therefore a
small equivalent width may imply a high global covering factor (e.g.,
Boroson \& Green 1992). Large column densities and global covering
factors may alleviate the current discrepancy between UV and X-ray
column densities.  Green et~al. (2001) found that LoBALs are weaker
X-ray sources than HiBALs, implying higher column densities for
LoBALs.  Other evidence that has been cited for high column densities
includes reddened optical and UV spectra (Sprayberry \& Foltz 1992;
Boroson \& Meyers 1992; Egami et~al.1996; Brotherton et~al. 2001) and
large Balmer decrements (Egami et~al.  1996).

As mentioned above, if the mass outflow rates are too high to be
maintained in steady--state, the outflows may originate from episodic
ejections, perhaps associated with the turning on of QSOs (Hazard
et~al. 1984).  LoBALs may be young quasars in the act of casting off
their cocoons of gas and dust (Voit et~al. 1993), related to
ultraluminous infrared galaxies (Egami 1999).  Becker et~al. (1997)
suggest that FeLoBALs may be the missing link between galaxies and
quasars.

It appears that for some QSOs the simple SYNOW code may be useful for
making rapid explorations and gaining some insight.  However, we have
merely characterized the LFR without explaining how it is heated, and
we have ignored the implication from the observed spectropolarization
of FeLoBALs (Brotherton et~al. 1997; Hutsem\'ekers, Lamy, \& Remy
1998; Schmidt \& Hines 1999; Lamy \& Hutsem\'ekers 2000) that not all
components can be spherically symmetric.  Detailed physically
self--consistent calculations with more powerful
synthetic--spectrum codes such as PHOENIX must be the basis for
ultimate decisions about the viability of the model sketched here.  We
intend to undertake such calculations for FeLoBALs.

\bigskip

We are grateful to Bob Becker for providing the spectrum of \1214 and
to Jules Halpern and an anonymous referee for constructive comments.
This work has been supported by National Science Foundation grants
AST--9986965 and AST--9731450 and by NASA grant NAGS--10171.

\clearpage

\begin {references}

\reference{} Becker, R. H., Gregg, M. D., Hook, I. M.,
McMahon,~R.~G., White,~R.~L., \& Helfand,~D.~J. 1997, ApJ, 479, L93

\reference{} Becker, R. H., White, R. L., Gree, M. D.,
Brotherton,~M.~S., Laurent--Muehleisen,~S.~A., \& Arav,~N. 2000, ApJ,
538, 72

\reference{} Boroson, T. A., 2002, ApJ, 565, 78

\reference{} Boroson, T., \& Green, R.., 1992, ApJS, 80, 109

\reference{} Boroson, T. A., \& Meyers, K. A., 1992, ApJ, 397, 442

\reference{} Branch, D., Baron, E., \& Jeffery, D. J., in Supernovae
and Gamma--Ray Bursts, ed. K.~W.~Weiler (New York: Springer--Verlag),
in press (2002); astro-ph/0111573

\reference{} Brotherton, M. S., Tran, H. D., van Breugel, W., Dey, A., \&
Antonucci, R., 1997, ApJL, 487, 113

\reference{} Brotherton, M., S., Tran, H. D., Becker, R. H., Gregg, M. D.,
Laurent-Muehliesen, S. A., \& White, R. L., 2001, ApJ, 546, 775

\reference{} de Kool, M., Becker, R. H., Gregg, M. D., White,~R.~L.,
\& Arav,~N.  2002, ApJ, 567, 58

\reference{} Drew, J., \& Giddings, J. 1982, MNRAS, 201, 27

\reference{} Egami, E., 1999, in Proc. "Galaxy Interactions at Low and High
Redshift", eds. J. E. Barnes and D. B. Sanders (IAU), 475

\reference{} Egami, E., Iwamuro, F., Maihara, T., Oya, S., \& Cowie, L. L., 1996,
AJ, 112, 73

\reference{} Ferland, G. J. 2000, Hazy, A Brief Introduction to Cloudy
94.00 (Univ. Kentucky)

\reference{} Graham, M. J., Clowes, R. G., \& Campusano, L. E. 1996,
MNRAS, 279, 1349

\reference{} Green, P. J., Aldcroft, T. L., Mathur, S., Wiles,~B.~J.,
\& Elvis,~M. 2001, ApJ, 558, 109

\reference{} Gregg, M. D., Becker, R. H., White, R. L.,
Richards,~G.~T., Chaffee,~F.~H., \& Fan,~X. 2002, ApJL, 573, L85

\reference{} Hamann, F., Korista, K. T., \& Morris, S. L., 1993, ApJ, 415, 541

\reference{} Hazard, C., Morton, D. C., Terlevich, R., \& McMahon, R., 1984, ApJ,
282, 33

\reference{} Hutsem\'ekers, D., Lamy, H., \& Remy, M., 1998, A\&A, 340, 371

\reference{} Kuraszkiewicz, J., Wilkes, B.\ J., Brandt, W.\ N., \& Vestergaard, M.,
2000, ApJ, 542, 631

\reference{} Kurucz, R. L. 1993, CD-ROM~1, Atomic Data for Opacity
Calculations (Cambridge: Smithsonian Astrophysical Observatory)

\reference{} Lamy, H., \& Hutsem\'ekers, D., 2000, A\&A, 356, L9

\reference{} Mathews, W. G., \& Ferland, G. J. 1987, ApJ, 323, 456

\reference{} Scargle, J. D., Caroff, L. J., \& Noerdlinger,
P. D. 1970, ApJ, 161, L115

\reference{} Schmidt, G. D., \& Hines, D. C., 1999, ApJ, 512, 125

\reference{} Sprayberry, D., \& Foltz, C. B. 1992, ApJ, 390, 39

\reference{} Surdej, J., \& Hutsem\'ekers, D. 1987, A\&A, 177, 42

\reference{} Surdej, J., \& Swings, J. P. 1981, A\&A, 96, 242

\reference{} Verner, E. M., Verner, D. A., Korista, K. T.,
Ferguson,~J.~W., Hamann,~F., \& Ferland,~G.~J. 1999, ApJS, 120, 101

\reference{} Voit, G. M., Weymann, R. J., \& Korsta, K. T. 1993, ApJ,
413, 95

\reference{} Vogt, S. S., et al. 1994, Proc. SPIE, 2198, 362

\reference{} Weymann, R. J., Morris, S. L., Foltz, C. B., \&
Hewitt,~P.~C. 1991, ApJ, 373, 23

\reference{} White, R. L. et al. 2000, ApJS, 126, 133

\end{references}

\begin{figure}
\plotone{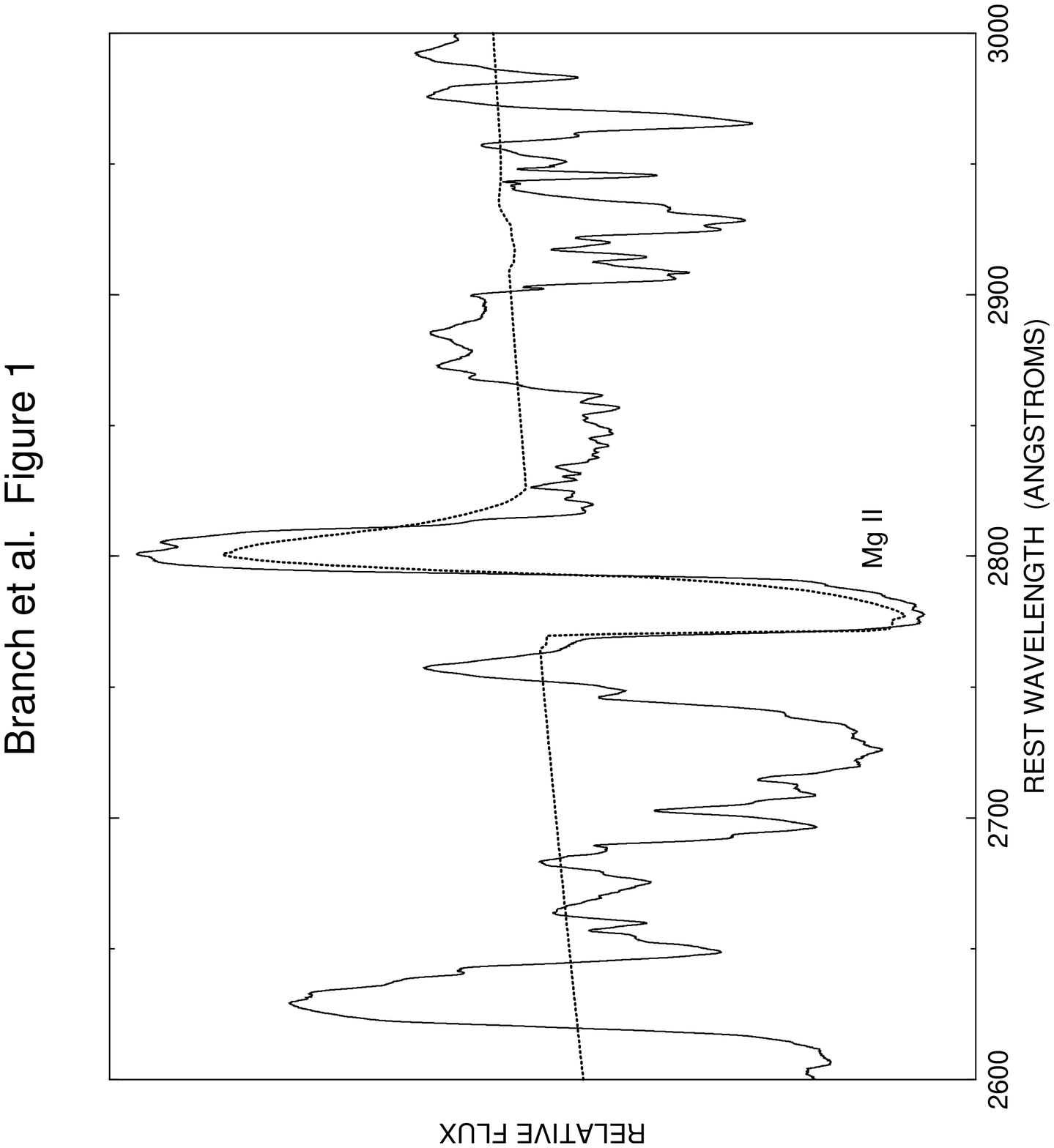}
\figcaption{The spectrum of \1214 in the region of the Mg~II $\lam2798$
feature ({\sl solid line}) is compared with a resonance--scattering
synthetic spectrum ({\sl dotted line}) that contains only lines of
Mg~II and has $v_{min}=1000$ \kms\ and $v_{max}=2800$ \kms.}

\end{figure}

\begin{figure}
\plotone{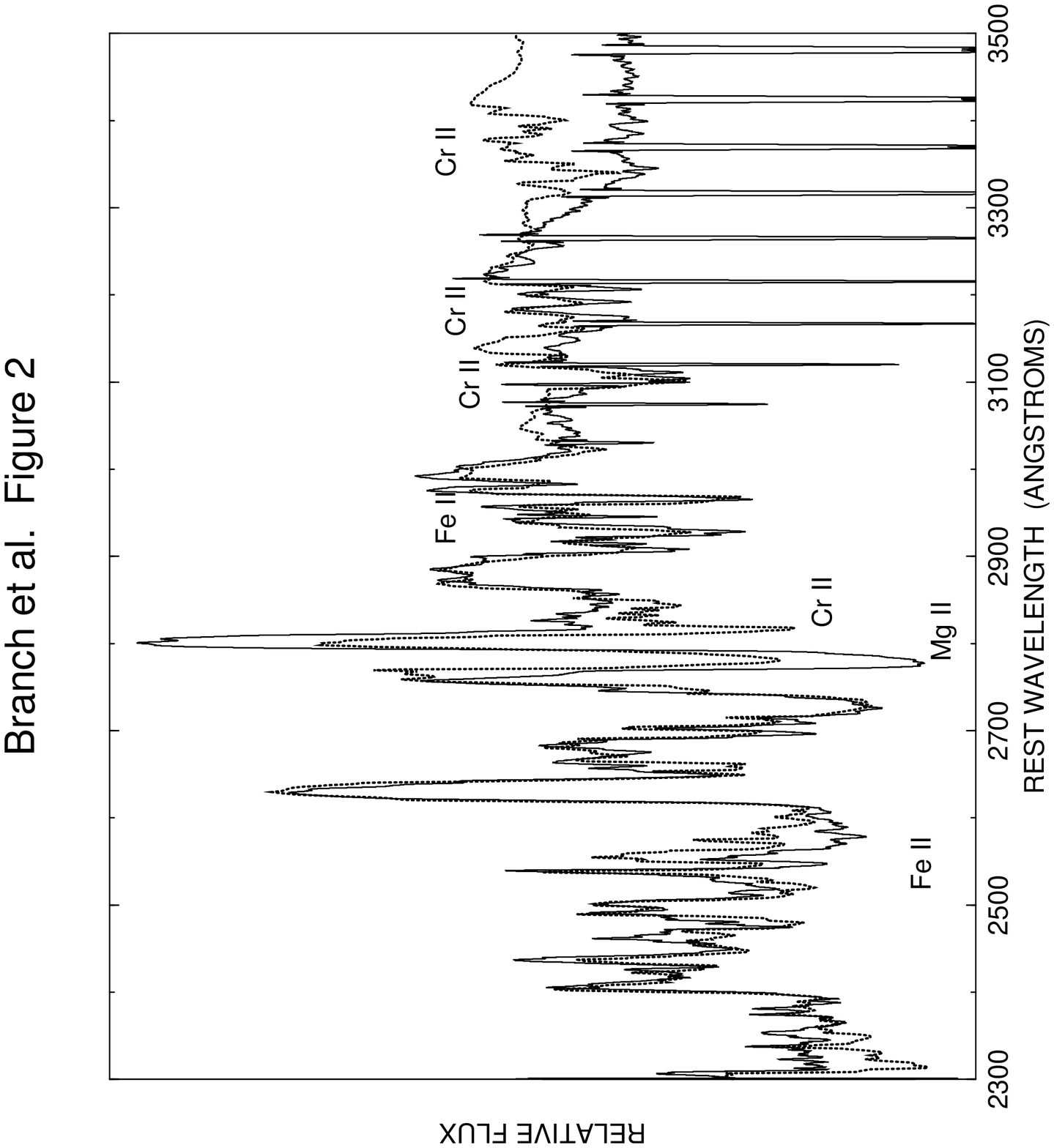}
\figcaption{The spectrum of \1214 ({\sl solid line}) is compared with a
resonance--scattering synthetic spectrum ({\sl dotted line}) that
contains lines of Mg~II having $v_{min}=1000$ \kms\ and $v_{max}=2800$
\kms, and lines of Fe~II and Cr~II having $v_{min}=1800$ \kms.  Fe~II
dominates the spectrum to the left of the Mg~II absorption.  The 10
narrow symmetric dips in the observed spectrum are gaps in the echelle
orders and have nothing to do with the spectrum of \1214.}

\end{figure}

\begin{figure}
\plotone{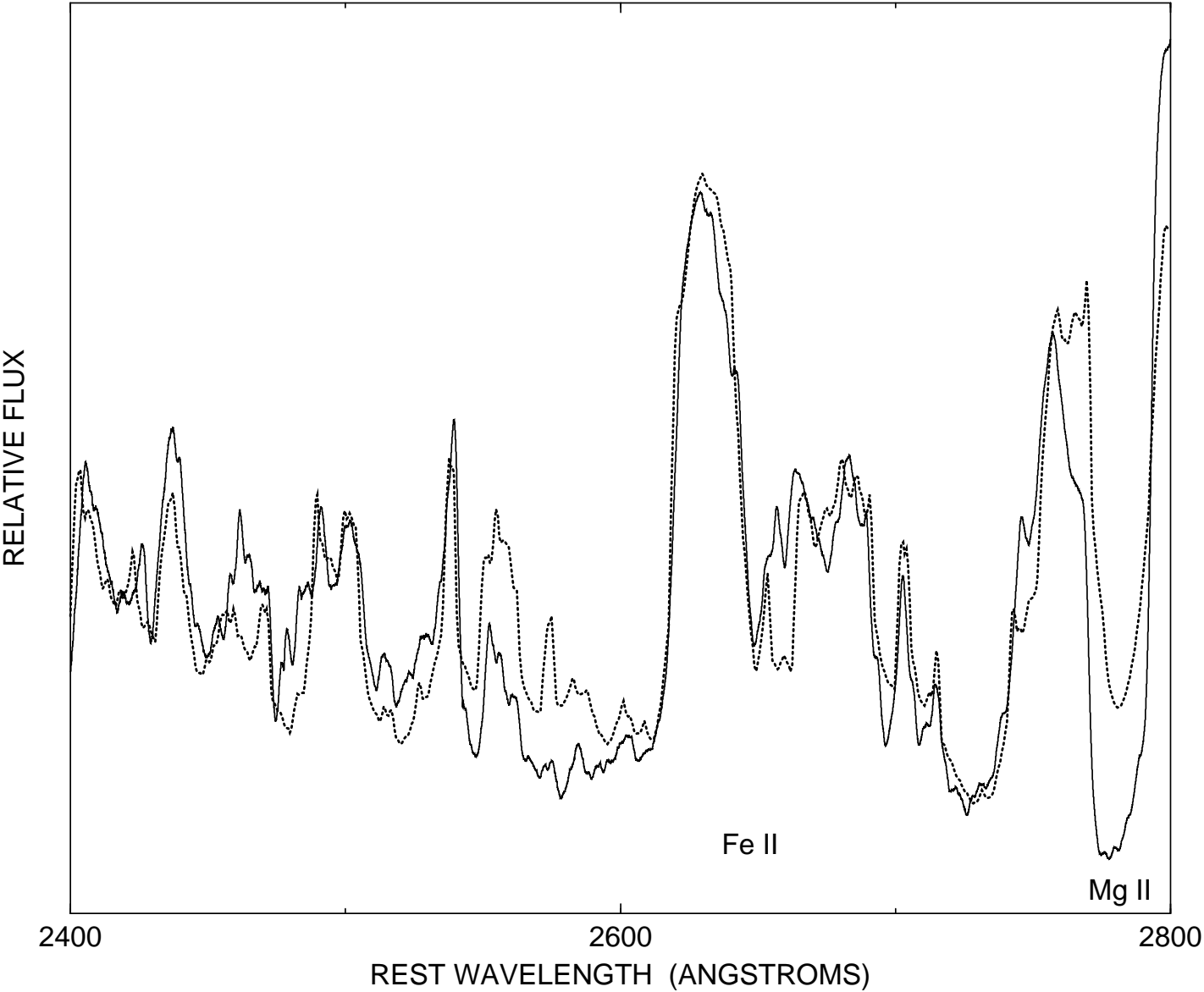}
\figcaption{A closer look at a crowded part of the observed and synthetic
spectra of Fig.~2.  Fe~II dominates the spectrum to the left of the
Mg~II absorption.}

\end{figure}

\end{document}